# Observation of tunneling gap in epitaxial ultrathin films of pyrite-type copper disulfide

Chong Liu(刘充)[1], Haohao Yang(杨好好)[1], Can-Li Song(宋灿立)[1,2], Wei Li (李渭)[1,2], Ke He(何珂)[1,2], Xu-Cun Ma(马旭村)[1,2], Lili Wang(王立莉)[1,2,†], Qi-Kun Xue(薛其坤)[1,2,†]

[1]*State Key Laboratory of Low-Dimensional Quantum Physics, Department of Physics, Tsinghua University, Beijing 100084*

[2]*Collaborative Innovation Center of Quantum Matter, Beijing 100084*

Correspondence to: liliwang@mail.tsinghua.edu.cn; qkxue@mail.tsinghua.edu.cn

We report scanning tunneling microscopy investigation on epitaxial ultrathin films of pyrite-type copper disulfide. Layer by layer growth of $CuS_2$ films with a preferential orientation of (111) on $SrTiO_3$(001) and $Bi_2Sr_2CaCu_2O_{8+\delta}$ substrates is achieved by molecular beam epitaxy growth. For ultrathin films on both kinds of substrates, we observed symmetric tunneling gap around Fermi level that persists up to ~ 15 K. The tunneling gap degrades with either increasing temperature or increasing thickness, suggesting new matter states at the extreme two dimensional limit.

Engineering of heterostructures and ultrathin films in the extreme of two dimensional (2D) limit is a vibrant frontier in realizing novel electronic states of matter. The most notable recent breakthrough includes the single-layer variants of various transition-metal dichalcogenides[1, 2] and the monolayer FeSe films grown on several $TiO_{2-\delta}$ terminated substrates[3-5], which exhibit exotic properties distinct from their bulk counterparts, therefore, have ignited explosive investigations. Aiming to search for new interfacial system or ultrathin films that exhibit diverse and novel properties, we chose $CuS_2$, a pyrite-type transition metal dichalcogenide, as new specimen.

The pyrite-type transition metal dichalcogenides $MX_2$, with M = Fe, Co, Ni, Cu, Zn and X = S, Se, Te (no $ZnTe_2$) show a large variety of electronic, magnetic and optical properties. Among the pyrites known so far, copper pyrites, $CuS_2$, $CuSe_2$ and $CuTe_2$, are the only superconducting compounds[6]. The pyrite-type $CuS_2$, which can be obtained by synthesis under high pressures, shows metallic behavior associated with a second-order phase transition at 150 K, weak ferromagnetism with a Curie temperature of 23 K[7, 8], and superconductivity at temperatures below $T_c$ = 1.5 K[9, 10]. Thus, there is possible coexistence of the superconductivity and the weak ferromagnetism below $T_c$. On the other hand, the previous transport, magnetization and specific heat measurements showed that there is no clear indication of strong electron correlations in the electronic properties of Cu pyrites[11]. It is chalcogen $p$ character rather than $d$ characteristic of Cu that dominates at the Fermi level[11, 12], which is in remarkable contrast to other 3$d$ transition metal pyrites.

To explore the diverse and novel properties in ultrathin films of $CuS_2$, in this work, we performed molecular beam epitaxy (MBE) growth and scanning tunneling microscopy/spectroscopy (STM/STS) study of ultrathin films of $CuS_2$ on $SrTiO_3$(001) and $Bi_2Sr_2CaCu_2O_{8+\delta}$ (Bi-2212) substrates. Under a moderate S-rich condition, the single-phase pyrite $CuS_2$ grows in a layer-by-layer mode with a preferential orientation of (111). At higher S flux, $CuS_2$ islands with various height instead of uniform films form, which conceives that S flux is a key factor for the growth of pyrite $CuS_2$. Another interesting finding is symmetric tunneling gap with weak coherence peaks around Fermi level, which occurs in ultrathin films and degrades with either increasing temperature or increasing thickness. The observation of such tunneling gap suggests new matter states in ultrathin films of $CuS_2$, which deserves further study.

Our experiments were conducted in a Createc ultrahigh vacuum STM system equipped with an MBE chamber for *in situ* sample preparation. The base pressure of the system is better than $1.0 \times 10^{-10}$ Torr. The samples were immediately transferred into the STM head for data collection at 4.6 K after MBE growth. A polycrystalline PtIr tip was used throughout the experiments. STM topographic images were acquired in a constant current mode, with the bias voltage ($V_s$) applied to the sample. Tunneling spectra were measured by disabling the feedback circuit, sweeping the sample voltage $V_s$, and then extracting the differential conductance d$I$/d$V$ using a standard lock-in technique with a small bias modulation of 673.2 Hz.

We chose $TiO_2$-terminated Nb-doped (0.05 wt. %) $SrTiO_3$(001) obtained after heating to 1100 ℃ and new cleaved optimal-doped Bi-2212 as substrates. Because of the extremely high vapor pressure of the S beam, the re-evaporation of S is a major concern especially in the precisely controlled growth of ultrathin $CuS_2$ films at the extreme 2D limit. In order to improve the sticking efficiency of S and the controllability of the chemical composition, we used CuS compound, which decomposes to solid $Cu_7S_4$ and $S_2$ vapor when heated[13], as a S source and set a comparatively low substrate temperature of 80 ℃. At such low substrate temperature, the surface migration of the arrived atoms would become less effective. In order to compensate this problem, we lowered the growth rate to ~ 0.05 monolayer (ML)/minute, accomplished by co-evaporating high-purity Cu (99.9999%) and CuS (99.98%) by Knudsen diffusion cells at cell temperatures of 850 ℃ and 170 ℃, respectively. Here, 1 ML refers to a nominal coverage which completely covers the $SrTiO_3$ or Bi-2212 surface. Under these conditions, the $CuS_2$ films grew in the layer-by layer mode and showed good crystal quality with a preferential surface orientation of (111), as will be shown later on.

Pyrite $CuS_2$ has a rock salt type structure, containing interpenetrating face-centered-cubic arrays of $Cu^{2+}$ cations and $S_2^{2-}$ anion dimers, with the lattice constant of 5.7898 Å[12]. Within a unit cell, the $Cu^{2+}$ cations occupy the face-centers of the cubic cell, while the $S_2^{2-}$ dimers are centered near the anion positions, as shown in upper left panel of Fig. 1(a). For the closest-packed (111) plane, as shown in right panel of Fig. 1(a), the surface exhibits a hexagonal structure, with an in-plane lattice constant of 8.19 Å, and each primitive unit cell (shown as a green rhombus) consists of four $Cu^{2+}$ cations and four $S_2^{2-}$ dimers. As schematically illustrated in bottom panel of Fig. 1(a), along the (111) direction, three of the four $S_2^{2-}$ dimers are tilted with respect to the (111) orientation, while the remaining $S_2^{2-}$ dimer is along the (111) direction. Therefore, the eight S atoms, separated into four different planes, are sandwiched between the two adjacent Cu layers with a distance of 3.34 Å. That is, the pyrite $CuS_2$ forms a quasi-layered structure along the (111) direction. For convenience, we hereafter introduce a reduced unit cell (including one Cu atom) marked by the red rhombus in upper right panel of Fig. 1(a). In the case of substrates, $TiO_2$ termination layer for $SrTiO_3$(001) and BiO termination layer for Bi-2212 have tetragonal structures with lattice constant of 3.9 Å and 3.8 Å, respectively.

Despite the structure mismatch between hexagonal in-plane structure of $CuS_2$ and tetragonal in-plane structures in either $SrTiO_3$(001) or Bi-2212 substrates, crystalline $CuS_2$ films with atomically flat (111) surface form on such two substrates. Figures 1(b)-(e) depict a series of typical STM topographic images of $CuS_2$ on $SrTiO_3$(001) at various coverages. Initially, at 0.7 ML, some irregular flat $CuS_2$ patches with uniform height distribute on the surface, as shown in Fig. 1(b). The apparent thickness of the $CuS_2$ patches is bias-dependent, increasing from 5.4 Å to 7.0 Å with sample bias decreasing from 4.5 V to 3 V, due to contrast between the density of states of $CuS_2$ and STO substrates. With coverage increased to 1 ML (Fig. 1(c)), the $CuS_2$ patches coalesce and form continuous films of 1 ML thick. Some second layer patches occur occasionally on the first layer $CuS_2$ films with constant thickness of 3.3 Å, exactly same as the layer distance along the (111) direction shown in Fig. 1(a), indicating a nearly layer-by layer growth. This layer-by-layer growth continues as the coverage increases. For example, at the nominal coverage of 7 ML, uniform and continuous film with atomically flat surface forms, except for small amount of redundant patches above that (Fig. 1 (d)). Shown in Fig. 1(e) is a zoom-in image taken on the flat terrace of 7 ML films. Although long-range ordered structure is hardly identified directly from topographic image, the corresponding fast Fourier transformation (FFT) image (inset of Fig. 1(e)) indicates hexagonal structure with lattice period of 4.0 Å, which is half the in-plane lattice constant of the CuS$_2$(111) plane and agrees excellently with the nearest Cu–Cu or S$_2$-S$_2$ distance depicted in upper right panel of Fig. 1(a). Considering that chalcogen *p* character rather than d characteristic of Cu dominantly contributes the density of states around the Fermi level[11], we attribute the atomic periodic structure observed here to S$_2^{2-}$ dimers. Combining the layer thicknesses and in-plane structure, we conclude that single-phase (111)-oriented pyrite CuS$_2$ thin films with S$_2$ termination form on SrTiO$_3$(001) substrate.

The most interesting finding in this work is the observation of symmetric tunneling gap around Fermi level. Figure 2 summarizes the differential tunneling spectra (d*I*/d*V*) taken on CuS$_2$ films of various thickness and at various temperatures. Displayed in Fig. 2(a) and (b) are d*I*/d*V* spectra of 1 ML CuS$_2$ taken at 4.6 K in a large bias range (-1.5 V – 2.0 V) and small bias range (-50 mV – 50 mV), respectively. We observed a band gap of 1.2 eV with the valence band top located at 0.4 eV above Fermi level (Fig. 2(a)). The gap value is consistent with band structure calculation result on bulk CuS$_2$, however, the gap position moves ~ 0.4 eV closer to Fermi level[12]. This shift suggests that the 1 ML CuS$_2$ films here are electron-doped compared with their bulk counterparts. This is most likely due to the charge transfer from the underlying SrTiO$_3$ substrates which act as charge reservoir, as previously revealed in FeSe/SrTiO$_3$[3-5] and LaAlO$_3$/SrTiO$_3$[14] hetero-structures. More striking is that a symmetric gap opens around Fermi level, characterized by weak coherence peaks at ± 10 meV and nonzero zero-bias-conductance (ZBC) (Fig. 2(b)). With increasing temperature up to 15 K, the gap degrades gradually in magnitude and ZBC increases simultaneously (Fig. 2(c)). This kind of tunneling gaps keep up for films at least thinner than 7 ML and exhibit thickness-dependent behaviors. For comparison, we summarized the typical d*I*/d*V* spectra of 1 ML, 3 ML and 7 ML CuS$_2$ films in Fig. 2(d). Clearly, the magnitude of gap decreases and the ZBC increases with increasing thickness, suggesting a low dimensional or interfacial origin of such tunneling gaps.

To modulate and further explore the properties of the CuS$_2$ films at 2D limit, we carried out the *in-situ* post-annealing of as-grown 1 ML CuS$_2$ films. As shown in Fig. 3(a), with annealing at 200 °C, the films evolve to striped-islands with heights increased to 1.6 – 2.0 nm, corresponding to 4 - 5 ML. Some of the striped-islands crossed SrTiO$_3$ steps, indicating the long-range diffusion of CuS$_2$. Such morphology change is accompanied by remarkable surface reconstruction and electronic properties change. Instead of 1×1 lattice, the surface exhibits a √3 ×√3-R30° reconstruction, as identified directly from the periodicity on the real space topographic image and the corresponding FFT image in Fig. 3(b). Figure 3(c) typifies the differential conductance d*I*/d*V* spectra of √3 ×√3-R30° surface. Compared with the 1×1 lattice, the semiconducting gap of 1.2 eV remains but shifts downwards and crosses the Fermi level. It seems that the CuS$_2$ films evolve to semiconducting states with the formation of √3 ×√3-R30° reconstruction from metallic states with 1×1 lattice. Our previous work on pyrite CuSe$_2$ showed that the chalcogenide concentration might lead to surface relaxation and thus various surface reconstructions[15], hence, we speculate that the above results observed on CuS$_2$ islands could be due to the S desorption with post-annealing. The desorption of S will lead to electron-doping, which agrees with the upward Fermi level shift shown in Fig. 3(c) compared with Fig. 2(a). It is worth noting that the spectra taken on the isolated CuS$_2$ islands may be affected by the possible non-ohmic contact between CuS$_2$ and SrTiO$_3$ substrates that the tunneling current passes through. To quantitatively character the difference between √3 ×√3 and 1×1 phases, comparing the spectra taken on the individual phases but in the same form of continuous films is required.

We also conducted annealing under S flux, and only found the films evolved to higher islands even under very low annealing temperature of 60 ℃ (Fig. 3(d)). This remarkable morphology variation, together with the Volmer–Weber growth at higher S flux (not shown), suggests that S flux plays an important role in the growth mode of $CuS_2$ films, which is in sharp contrast with the growth of other chalcogenides such as selenides[3-5, 15]. $CuS_2$ islands prepared under higher S flux show similar spectral features as those formed after annealing under S flux, which again indicates the susceptibility of electronic properties to S. This extra susceptibility naturally results in spatially inhomogeneous properties, as we observed spatially inhomogeneous spectra at each states. Here, we mainly show the most characteristic and reproducible ones.

To identify the tunneling gap shown in Fig. 2 is mainly due to intrinsic low dimensional or extrinsic interfacial effect, we grew $CuS_2$ films on another substrate — the BiO surface of the new cleaved Bi-2212 crystals. As displayed in Fig. 4(a), the BiO surface shows the characteristic b-axis supermodulation with a period of 26 Å and square lattice with a lattice constant of 3.8 Å. On such BiO surface, the $CuS_2$ grows via the same layer-by-layer mode and the step height is 3.3 Å (Fig. 4b). On the surface of 1 ML $CuS_2$, as shown in Fig. 4c, the b-axis supermodulation of BiO surface still can be seen clearly. Despite that, the FFT image shows hexagonal lattice with periodic constants of 6.9 Å, which agrees well with that of $\sqrt{3} \times \sqrt{3}$-R30° reconstruction. Similar to $CuS_2$ films on $SrTiO_3(001)$ substrates, the Fermi level lies at ~ 0.4 eV below the valence band top (Fig. 4(e)) and a tunneling gap with weak coherence peaks at ±10 meV occurs around the Fermi level (Fig. 4(f)). On the other hand, in the exposed BiO regions, the surface structure looks same as the new-cleaved BiO surface and the differential conductance d$I$/d$V$ spectra also show the characterized pseudo-gap of 45 meV (Fig. 4(d)), which hints sharp interface between $CuS_2$ films and BiO layer.

We compare the morphologies and spectra of $CuS_2$ films on $SrTiO_3(001)$ and Bi-2212 substrates and summarize our results as below. (1) The band gap position, as identified from the spectra shown in Figs. 2(a) and 4(e), moves ~ 0.4 eV closer to Fermi level compared with the calculation results for bulk $CuS_2$[12]. This suggests that both $SrTiO_3$ and Bi-2212 substrates could introduce electron doping to $CuS_2$ films, while low dimensional effect could also contribute to LDOS change. (2) Despite various interfaces, symmetric tunneling gap with magnitude of ~ 10 meV can be consistently observed on both $SrTiO_3(001)$ and Bi-2212 substrates, which suggests intrinsic novel properties in the ultrathin films of $CuS_2$. (3) The tunneling gap degrades with either increasing thickness or increasing temperature, suggesting new matter states in ultrathin $CuS_2$ films that become strongest at the extreme 2D limit. We didn't see signature of charge density waves. In contrast to Coulomb gap previously observed on semiconducting films[16], here the relatively large tunneling gap is observed on metallic state. Although no noticeable strong electron correlation has been shown in its bulk counterparts[11], the ultrathin films of $CuS_2$ with interfacial doping could probably exhibit exotic properties. The tunneling gap observed on $CuS_2$ films exhibit similar features to the pseudo-gap in cuprate, as clearly seen from the comparison between the spectra shown in Figs. 2(b)-(d), 4(f) and the spectrum taken on BiO surface shown in Fig. 4(d). The origin of the tunneling gap deserves further study. (4) The contrasting morphologies and surface reconstructions, i.e. films vs. islands and metallic 1×1 vs. semiconducting $\sqrt{3} \times \sqrt{3}$, can be both tuned by S flux.

Similarly, the electronic properties are susceptible to S flux, which is a significant feature of sulfide compound. This requires fine controllability on the sample stoichiometry in study on such sulfide compounds.

In summary, we successfully prepared single crystalline films of $CuS_2$ by using MBE and observed tunneling gap with weak coherence peaks in ultrathin films at the extreme 2D limit. The method of MBE preparation of ultrathin $CuS_2$ films could be used for other transition metal disulfides, such as $MoS_2$ and $TaS_2$, which exhibit diverse interesting features of optical and electrical properties and are being hotly studied. Compared with exfoliation and chemical vapor deposition methods that previously used, the MBE growth guarantees single phase crystalline films with fine stoichiometry controllability. The symmetric tunneling gap with weak coherence peaks around Fermi level suggests new matter states in ultrathin films of $CuS_2$, which deserves further study.

**Acknowledgements**

The work is supported by the National Natural Science Foundation of China under Grant Nos 11574174, 11774193 and 11790311, and the National Basic Research Program of China under Grant No 2015CB921000.

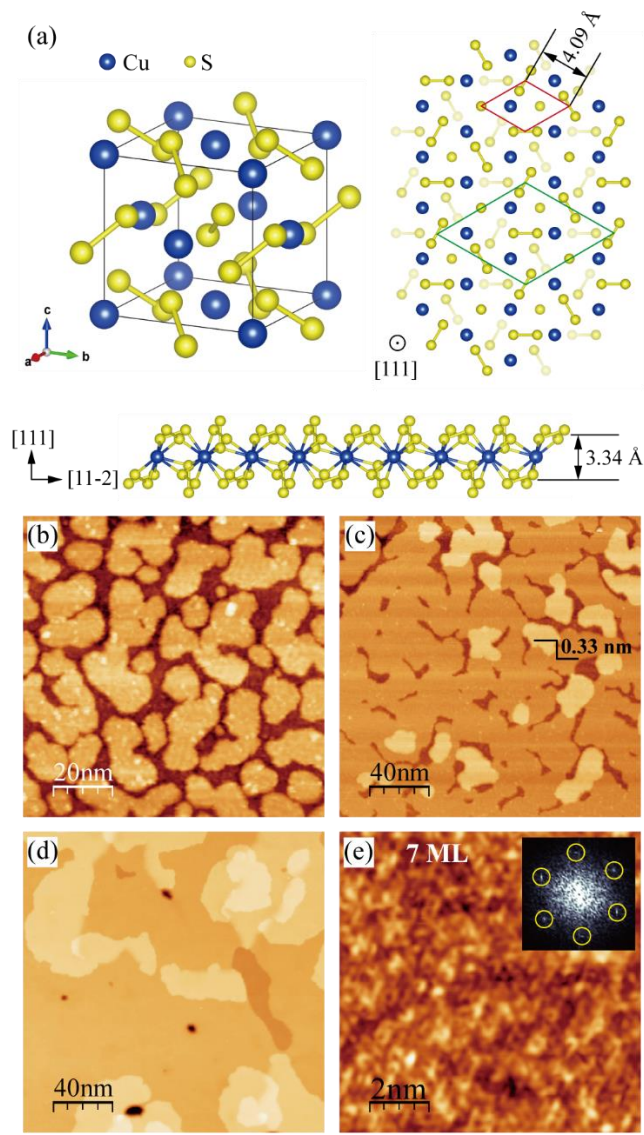

Fig. 1.  (a) Crystal structure of pyrite $CuS_2$. (b)-(e) STM topographic images of $CuS_2$ on $SrTiO_3$(001) at various nominal coverages: (b) 0.7 ML ($V_s$ = 5 V, $I$ = 50 pA), (c) 1 ML ($V_s$ = 4.5 V, $I$ = 40 pA), and (d)-(e) 7 ML ((d) $V_s$ = 6 V, $I$ = 50 pA and (e) $V_s$ = 0.4 V, $I$ = 100 pA). The inset in (e) is the corresponding FFT images with the yellow circles marking the 1×1 Bragg points.

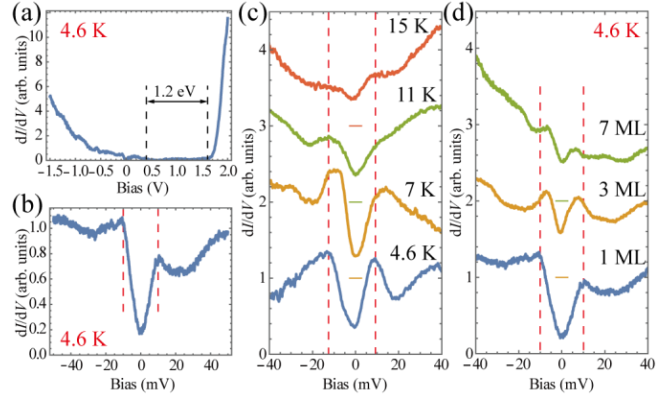

**Fig. 2.** (a) and (b) d$I$/d$V$ tunneling spectra taken on 1 ML CuS$_2$ films on SrTiO$_3$(001) ((a) $V_s$ = 2 V, $I$ = 100 pA; (b) $V_s$ = -50 mV, $I$ = 100 pA). (c) Tunneling spectra taken on 1 ML films at various temperatures ($V_s$ = 60 mV, $I$ = 100 pA). (d) Tunneling spectra taken on films with various thickness: 1 ML ($V_s$ = -50 mV, $I$ = 100 pA), 3 ML ($V_s$ = -60 mV, $I$ = 200 pA) and 7 ML ($V_s$ = 60 mV, $I$ = 100 pA). The spectra in (c) and (d) are shifted along the vertical axis. The horizontal bars indicate the zero-conductance position of each curve. The black dashed lines in (a) show the band gap. The red dashed lines in (b)-(d) are guide for eyes, showing the change of tunneling gap.

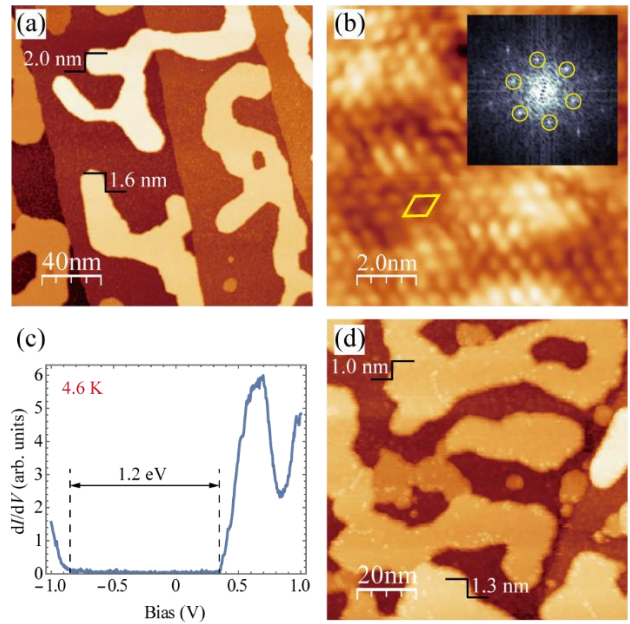

**Fig. 3.** (a)-(b) Topographic image of 1 ML CuS$_2$ after being annealed at 200 ℃ ((a) $V_s$ = 2.5 V, $I$ = 50 pA; (b) $V_s$ = 1.3 V, $I$ = 200 pA). In (b), the yellow rhombus labels the $\sqrt{3} \times \sqrt{3}$ unit cell, and the inset is the corresponding FFT images showing $\sqrt{3} \times \sqrt{3}$ reconstruction, as labelled by the yellow circles. (c) d$I$/d$V$ tunneling spectra taken on CuS$_2$ islands on SrTiO$_3$(001) ($V_s$ = 1 V, $I$ = 100 pA). The black dashed lines show the band gap. (d) Topographic image of 1 ML CuS$_2$ after being annealed under S flux at 60 ℃ ($V_s$ = 3 V, $I$ = 50 pA).

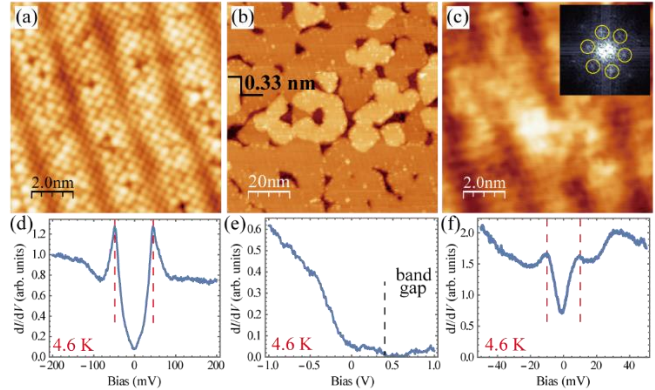

**Fig. 4.** (a) Topographic image of as-cleaved Bi-2212 ($V_s = 0.6$ V, $I = 100$ pA). (b) and (c) Topographic images of CuS$_2$ grown on Bi-2212 with nominal coverage of 1 ML ((b) $V_s = 3$ V, $I = 50$ pA; (c) $V_s = 0.5$ V, $I = 100$ pA). The inset of (c) is the FFT image showing $\sqrt{3} \times \sqrt{3}$ reconstruction as labeled by the yellow circles. (d) Tunneling spectrum taken on exposed Bi-2212 after th4be growth of CuS$_2$ ($V_s = -0.2$ V, $I = 100$ pA). (e) and (f) Tunneling spectra taken on 1 ML CuS$_2$ on Bi-2212 ((e) $V_s = -1$ V, $I = 50$ pA; (f) $V_s = -50$ mV, $I = 100$ pA). The red dashed lines in (d) and (f) show coherence peaks of the tunneling gaps.